\pdfoutput=1
\documentclass[sigconf]{acmart}
\AtBeginDocument{%
  \providecommand\BibTeX{{%
    \normalfont B\kern-0.5em{\scshape i\kern-0.25em b}\kern-0.8em\TeX}}}

\copyrightyear{2020}
\acmYear{2020}
\setcopyright{iw3c2w3}
\acmConference[WWW '20]{Proceedings of The Web Conference 2020}{April 20--24, 2020}{Taipei, Taiwan}
\acmBooktitle{Proceedings of The Web Conference 2020 (WWW '20), April 20--24, 2020, Taipei, Taiwan}
\acmPrice{}
\acmDOI{10.1145/3366423.3380169}
\acmISBN{978-1-4503-7023-3/20/04}

\usepackage{url}
\usepackage{amsmath}
\usepackage{color}
\usepackage{xspace}
\usepackage{caption}
\usepackage{graphicx}
\DeclareGraphicsExtensions{.pdf}
\usepackage{multirow}
\usepackage{subfig}
\usepackage{listings}
\usepackage{flushend}
\usepackage{pifont}
\usepackage{balance}
\usepackage{hyperref}
\usepackage{amssymb}
\usepackage{capt-of}
\usepackage{gensymb}
\usepackage{balance}
\usepackage{xcolor}
\usepackage{wrapfig}
\usepackage{picins}

\hypersetup{draft} 
\newcommand{\etc}{\textit{etc.}\xspace}
\newcommand{\ie}{\textit{i.e.,}\xspace}
\newcommand{\eg}{\textit{e.g.,}\xspace}

\newcommand{\hd}[1]{\small{\textbf{\texttt{#1}}}\normalsize}

\newcommand{\hdf}[1]{{\textbf{\texttt{#1}}}\footnotesize}

\newcommand{\BULLET}{\vspace{+.05in} \noindent $\bullet$ \hspace{+.00in}}

\newcommand{\squishlist}{
	\begin{itemize}[noitemsep,nolistsep]
		\setlength{\itemsep}{-0pt}
	}
	\newcommand{\squishend}{
	\end{itemize}
}

\newcommand{\DEG}{\degree\xspace}
\newcommand{\feng}[1]{#1}

\newcommand{\arvind}[1]{\textcolor{black}{#1}}

\newcommand{\vz}{\hd{VZ}\xspace}
\newcommand{\vzf}{\hdf{VZ}\xspace}

\newcommand{\mysection}[1]{\vspace{-.02in}\section{#1}\vspace{-.00in}}
\newcommand{\mysubsection}[1]{\vspace{-.02in}\subsection{#1}\vspace{-.00in}}

\begin{document}

\title{A First Look at Commercial 5G Performance on Smartphones}


\author{Arvind Narayanan, Eman Ramadan, Jason Carpenter, Qingxu Liu, Yu Liu, Feng Qian, Zhi-Li Zhang}
 \affiliation{%
   \institution{University of Minnesota -- Twin Cities}
 }
 \email{fivegophers@umn.edu}


\begin{abstract}
We conduct to our knowledge a first measurement study of commercial 5G
performance on smartphones by closely examining 5G networks of three carriers (two mmWave carriers, one mid-band carrier) in three U.S. cities.
We conduct extensive field tests on 5G performance in diverse urban environments.
%
We systematically analyze the handoff mechanisms in 5G and their impact on network performance.
We explore the feasibility of using
location and possibly other environmental information to predict the network performance.
We also study the app performance (web browsing and \texttt{HTTP} download) over 5G.
%
Our study consumes more than 15 TB of cellular data.
Conducted when 5G just made its debut, it
provides a ``baseline'' for studying how 5G performance evolves, and identifies key research directions on improving 5G users' experience in a cross-layer manner.
\arvind{We have released the data collected from our study (referred to as \emph{5Gophers}) at
\url{https://fivegophers.umn.edu/www20}.}

\end{abstract}



\maketitle

\mysection{Introduction}


2019 marks the year for 5G, which was eventually rolled out for commercial services to consumers.
Compared to 4G LTE, 5G is expected to offer significantly higher bandwidth, lower latency, and better scalability (\ie supporting more devices).
The mainstream 5G deployment employs the millimeter wave (mmWave) technology that
can provide, in theory,  a throughput of up to 20 Gbps -- a 100$\times$ improvement compared to today's 4G~\cite{qualcomm}.\footnote{\feng{In this paper, we use ``4G'' to refer to 4G LTE networks.}}
Under the hood, this is achieved by a series of innovations including
massive MIMO, advanced channel coding, and scalable modulation.
%

5G is expected to fuel a wide range of applications that cannot be well supported by 4G, such as ultra-HD (UHD) video streaming, networked VR/AR, low-latency cloud gaming, and vehicle-to-everything (V2X) communication. Despite these potentials,
commercial 5G services are at their infancy. In early summer 2019, \hd{Verizon} (\vz) launched 5G in Chicago and Minneapolis. It uses a 400 MHz channel at 28 GHz, making it the world's first commercial mmWave 5G service for consumers. 
Followed by that, three other major U.S. carriers (\hd{T-Mobile}, \hd{Sprint}, and \hd{AT\&T}) have also rolled out their 5G services (Table~\ref{tab:carriers}).
Many major carriers around the world are in the process of commercializing 5G.

\begin{table}[t]
\vspace{1em}
\small
\feng{
\begin{tabular}{|c||c|c|c|c|}
\hline
\textbf{Carrier}    & \textbf{Verizon}~\cite{vz_5g}  & \textbf{AT\&T}~\cite{att_5g}     & \textbf{T-Mobile}~\cite{tmobile_5g}       & \textbf{Sprint}~\cite{sprint_5g} \\ \hline
\textbf{Type}       & mmWave         & mmWave         & \arvind{mmWave}         & mid-band \\
\textbf{Freq.}  & 28/39 GHz & 24/39 GHz & 28/39 GHz & 2.5 GHz \\
\hline
\end{tabular}%
}
\vspace{1em}
\caption[ABC]{5G technologies adopted by major U.S. carriers.\footnotemark}
\vspace{-.3in}
\label{tab:carriers}
\end{table}
\footnotetext{\arvind{Based on the 5G services provided by the carriers as of October 2019.}}

Commercial mmWave 5G operates at a much higher frequency from 24 GHz to 53 GHz with abundant free spectrum. On the positive side, this offers much higher bandwidth compared to 4G.
%
%
On the negative side, due to its short wavelength, mmWave signals propagate in a pseudo-optical manner, and are vulnerable to attenuation and blockage.
%
mmWave has been studied theoretically and experimentally over various testbeds as a standalone physical layer (\S\ref{sec:related}).
%
However, integrating mmWave into commercial 5G networks faces far more challenges than those on the physical layer itself. Indeed, 5G is a complex ecosystem involving multiple entities -- UEs (user equipments, \ie the client device), base stations, the core network, and even the Internet needs to embrace the 5G as our study will demonstrate; 5G also interacts with multiple protocol layers, with numerous optimization opportunities even at the transport and application layers;
furthermore, 5G needs to properly coexist with the legacy 4G for a very long time.


In this paper, we conduct to our knowledge a first measurement study of commercial mmWave 5G networks on commodity smartphones. This study presents several challenges.

\BULLET The breadth of the 5G ecosystem makes it non-trivial to determine \emph{what to study}.
To this end, we take a user-centric approach by strategically selecting the measurement subjects that incur a high impact on end users' experience: transport-layer performance under varying mobility scenarios and the application Quality-of-Experience (QoE). Given the importance of mmWave, we pay particular attention to the implications of mmWave in our measurements.

\BULLET Due to 5G's very recent debut, there does not exist mature tools for capturing and monitoring 5G-specific information such as service status and 5G/4G handoff events. We therefore make methodological contributions by developing our own measurement tools.

\BULLET mmWave's sensitivity to the environment dictates us to examine a wide range of factors that can influence the mmWave performance, complicating our measurement. We thus carefully select key factors including the UE-tower distance, UE orientation, blockage, and the weather; we then strategically design the controlled experiments to study their performance impact.

\BULLET mmWave's ultra-high bandwidth makes it much more likely that the Internet becomes the performance bottleneck -- a problem seldom appearing in 4G. We take two approaches to tackle this challenge. For most of the experiments, we conduct various pilot studies to maximize our confidence that the Internet does not remain the bottleneck; meanwhile, when the Internet-side bottleneck is inevitable (\eg the latency), we also experimentally reveal how the Internet part can affect the 5G performance in realistic settings.

\BULLET Last but not least, there also exist non-technical obstacles. For example, our team needs to travel to multiple cities to conduct field measurements for multiple 5G carriers.

We study three major 5G carriers in this work: \hd{VZ}, \hd{T-Mobile}, and \hd{Sprint}\footnote{\feng{When we wrote this paper, \hdf{AT\&T} has not yet offered 5G to non-business customers.}} in three U.S. cities: Minneapolis, Chicago, and Atlanta. Among them, \vz and \hd{T-Mobile} employ the mmWave technology, while \hd{Sprint} adopts the ``mid-band'' 5G operating at 2.5 GHz at which the signal propagation still remains omni-directional. We conduct all experiments on commercial smartphones. Overall, our study consists of the following aspects.

\textbf{An Overview of Today's 5G Performance (\S\ref{sec:landscape}).}
We begin by providing a first impression on the performance of today's commercial 5G services by
measuring the throughput and latency of the three carriers' 5G networks.
The results indicate that under typical urban environments where the UE is stationary, the average mmWave 5G throughput significantly outperforms that of the mid-band 5G.
However, today's commercial 5G offers little latency improvement due to its Non-Standalone (NSA) Deployment model that shares much of the existing 4G infrastructure with 5G (\S\ref{sec:related}).

\textbf{5G Performance of Stationary UE (\S\ref{sec:sta}).}
We compare the performance of mmWave 5G and 4G on key metrics such as throughput, latency, and packet loss rate when the UE remains stationary.
We conduct our experiments under diverse scenarios with different distances/orientations between the 
UE and the 5G-NR (New Radio) panel, obstruction levels, and weather conditions.
We find that commercial mmWave 5G offers much higher throughput than 4G ($\sim$10x improvement). However, even under clear line-of-sight, 5G throughput exhibits much higher variation than 4G, mainly due to the PHY-layer nature of mmWave signals (\S\ref{sec:los}).
Under non-line-of-sight (NLoS), mmWave 5G signals can be easily blocked by hands or human body. Despite that, in realistic urban environments, surrounding signal reflections can oftentimes mitigate the performance degradation, allowing 5G to function under NLoS (\S\ref{sec:nlos}).

\textbf{Mobility Performance (\S\ref{sec:hand}).} We investigate the mmWave performance when the UE is moving (\eg a 5G user walking or driving). We find that 4G-5G handoffs can be triggered frequently
by either network condition change or user traffic.
Even under low mobility (\eg walking), a smartphone may experience 30+ 4G/5G handoffs in less than 8 minutes.
Such a large number of switches
may confuse applications (\eg the video bitrate adaptation logic) and bring highly inconsistent user experiences. Compared to mmWave 5G, mid-band 5G offers better mobility performance due to its omni-directional radio. For the same reason, 4G also exhibits much better stability when the UE is moving.
These results indicate the necessity to jointly utilize mmWave 5G and omni-directional radio such as 4G in mobility scenarios (\eg through MPTCP~\cite{nikravesh2016depth}) where 4G can help guarantee the basic data connectivity.

\textbf{Inefficiency of Location-based Performance Estimation \linebreak (\S\ref{sec:loc}).}
In 3G/4G, location is known to be useful for predicting the cellular performance~\cite{schulman2010bartendr,margolies2016exploiting}. We investigate the feasibility of performing
location-based performance prediction in mmWave 5G through a 30-day field study, and
find that at a given location, mmWave 5G exhibits a statistically higher throughput variation compared to 4G,
due to mmWave's sensitivity to the environment -- a small perturbation
can affect the performance, making the location-based throughput prediction difficult.

\textbf{Application Performance (\S\ref{sec:app_perf}).} We study real application performance over mmWave 5G.
We find that for web browsing, today's 5G brings benefits only for large web pages; meanwhile, the optimizations brought by \texttt{HTTP/2} and \texttt{HTTP/3} (\texttt{QUIC}) are effective over 5G (\S\ref{sec:web}).
%
For large \texttt{HTTP(S)} download, we make an interesting finding that the goodput is significantly lower than the available mmWave 5G bandwidth, because many cross-layer factors may slow down the download. For example, compared to \texttt{HTTP}, \texttt{HTTPS} increases the average median download time by 23.5\% due to the \texttt{TLS} overhead (\S\ref{sec:dl}).
%
Overall, our results indicate that mmWave
5G's high bandwidth does not always translate to a better application QoE, whose improvement requires joint, cross-layer optimizations from multiple players in the mobile ecosystem.

We make the following contributions in this paper.

\BULLET We develop practical and sound measurement methodologies for 5G networks on COTS smartphones.

\BULLET We present timely measurement findings of mmWave 5G and mid-band 5G performance on smartphones with key insights.
Our experiments constitute more than 15 TB traffic\footnote{We purchased multiple unlimited 5G data plans from \hdf{VZ} , \hdf{T-Mobile}, and \hdf{Sprint} for this study. Our study conforms to all the carriers' wireless customer agreements.} and span three major 5G carriers in the U.S.
As they were conducted when commercial 5G had just made its debut, we expect our results to provide an important ``baseline'' for studying how 5G performance evolves.

\BULLET We \arvind{release our measurement dataset}, referred to as \emph{5Gophers}, to the research community to benefit work that needs real 5G data. The URL of the dataset is:

\vspace{+.05in}
\begin{center}
  \url{https://fivegophers.umn.edu/www20}
\end{center}

\mysection{Background and Related Work}
\label{sec:related}

\textbf{mmWave} is an innovative technology integrated into 5G. Unlike 3G/4G that works at $\le$5 GHz, mmWave 5G radios operate at much higher frequencies
of 24 to 53 GHz (according to 3GPP 38.101~\cite{3gpp_38_101}) with considerably abundant free spectrum.
Despite its high bandwidth,
mmWave's short
wavelength makes its signals vulnerable to attenuation. To overcome this, mmWave transceivers have to use phased-array antennas to form highly directional beams. Due to the pseudo-optical nature of a beam, the signals are sensitive to blockages such as a pedestrian or a moving vehicle.
Switching from line-of-sight (LoS) to \arvind{non-line-of-sight} (NLoS) due to blockage may cause significant data rate drop or even complete blackout despite the beamforming algorithm that attempts to ``recalibrate'' the beams by seeking for a reflective NLoS path~\cite{nitsche2015steering, sur2017wifi}.

Researchers have demonstrated the feasibility of deploying
mmWave in data centers~\cite{halperin2011augmenting, zhou2012mirror,zhu2014cutting},
indoor~\cite{xu2002spatial, anderson2004building, sur201560, tie201260, collonge2004influence, nitsche2015boon, haider2018listeer, abari2017enabling}, and outdoor environments~\cite{zhao201328, rappaport2013millimeter, zhu2014demystifying, rappaport2013broadband, rappaport201238, qiu2018avr}, as well as have conducted studies on beamforming and beam tracking~\cite{giordani2016mmwave,palacios2017mmwave,roh2014mmwave}.
But none of them studies mmWave in commercial 5G context on smartphones.

\feng{
\textbf{Mid-band 5G.} Instead of adopting the mmWave technology, some carriers deploy their 5G networks over the mid-band frequency (1--6 GHz) whose
radio signal largely remains omni-directional and offers a decent data rate. Mid-band 5G forms the basis of initial 5G services, but may suffer from a lack of spectrum in the long term. In contrast, mmWave has the unique advantages of ultra-high speed and abundant spectrum despite its limitations such as high attenuation and pseudo-optical signal propagation~\cite{gsma_5g_spectrum}. As shown in Table~\ref{tab:carriers}, three out of the four major carriers in the U.S. have adopted mmWave as the 5G technology.
}

\textbf{5G Infrastructure.}
To reduce the time to market, carriers couple their 5G core network equipment with existing 4G LTE infrastructure in what is known as \textit{Non-Standalone Deployment} (NSA). 
NSA utilizes 5G-NR for data plane operations while retaining their 4G infrastructure for control plane operations~\cite{erikkson}.
NSA is contrasted with \textit{Standalone Deployment} (SA) -- fully independent of legacy cellular infrastructures.
\feng{All carriers in Table~\ref{tab:carriers} currently employ the NSA model for their first commercial 5G deployment.}

\mysection{Measurement Methodology}
\label{sec:setup}

\begin{figure}[t]
    \centering
    \includegraphics[width=0.43\textwidth]{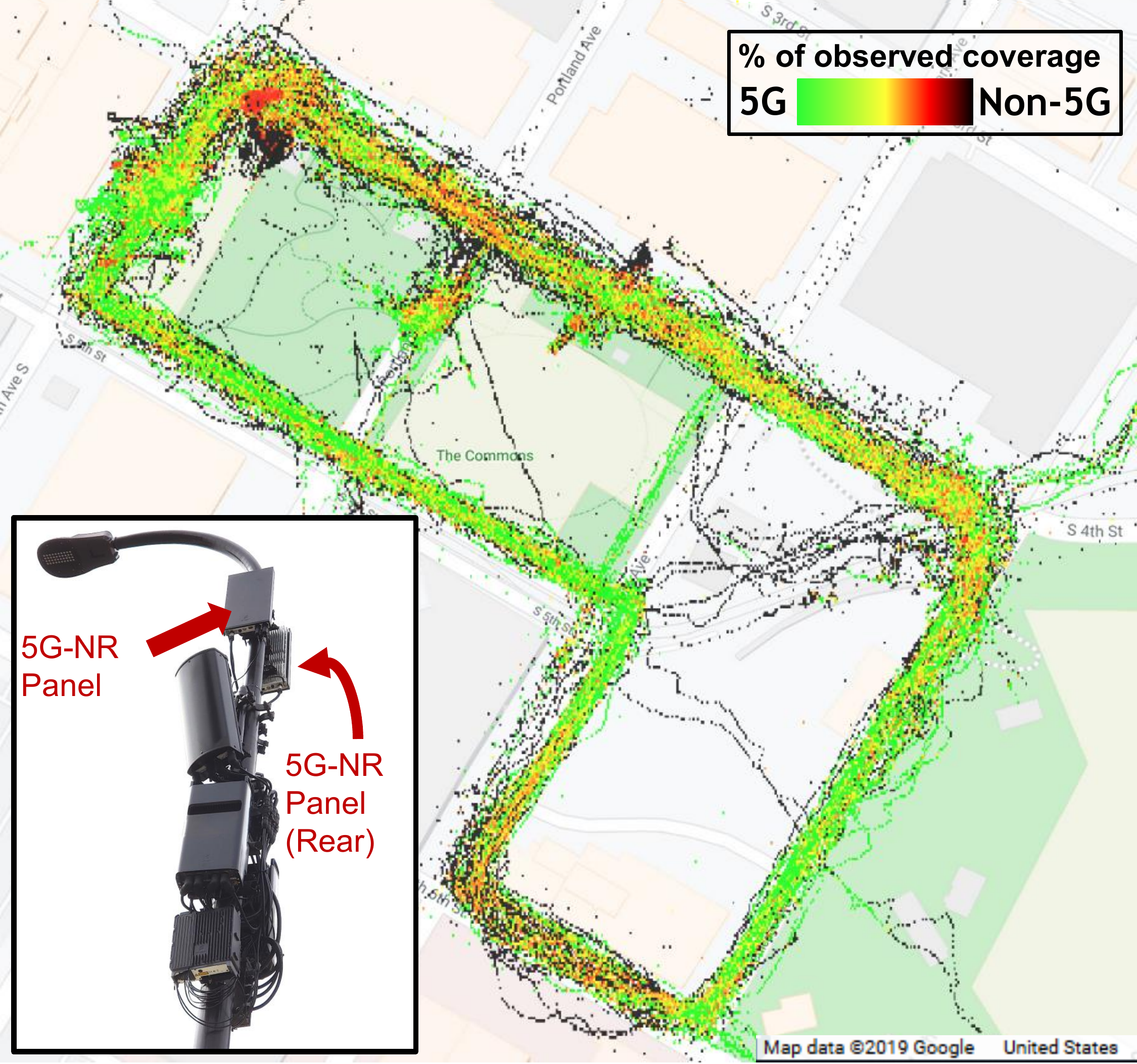}
    \vspace{-.12in}
    \caption{5G coverage recorded at Minneapolis's Commons Park.
    A color gradient from green to black indicates the percentage of observed 5G coverage (high to low respectively). We sampled 6.8 million data points to inform this visualization. Also indicated is a 5G mmWave base station.}
    \vspace{-.23in}
    \label{fig:coverage}
\end{figure}

\textbf{5G Networks.}
\feng{Most of} our experiments were conducted over \vz's 5G network.
\feng{In summer 2019 when we started this study,} \vz was the only cellular carrier in the U.S. that offers commercial mmWave-based 5G services to consumers at specific downtown areas in two cities: Minneapolis and Chicago. 
%
In their 5G coverage areas, dense 5G base stations are deployed (\eg about 10 within two blocks in downtown Chicago). Due to \vz's adoption of NSA (\S\ref{sec:related}), 5G base stations are typically co-located with or very close to those of 4G (based on our knowledge and visual inspection).
\feng{As shown in Figure~\ref{fig:coverage},}
a 5G mmWave base station is typically equipped with one or more \emph{panels} that are the mmWave transceivers.
We observe that the panels typically face populated areas such as streets and pedestrian walkways.
%
Figure~\ref{fig:coverage} also exemplifies the 5G coverage in a downtown area of Minneapolis, based on 6.8 million data points collected from our 4-month field study.

In addition to \vz, for comprehensiveness, we also study two other carriers (\hd{T-Mobile} and \hd{Sprint}) listed in Table~\ref{tab:carriers}. \hd{T-Mobile} also uses mmWave and \hd{Sprint} employs a mid-band frequency at 2.5 GHz. We experimentally study their performance in~\S\ref{sec:landscape}.

\textbf{5G User Equipment (UE).}
We use two types of COTS 5G-capable smartphones: Motorola Moto Z3 and Samsung Galaxy S10 5G (SM-G977U), henceforth referred to as MZ3 and SGS10, respectively.
%
SGS10 has a built-in 5G radio, while MZ3 requires a separate accessory called 5G Mod~\cite{mz3_5gmod} for accessing 5G.
Comparing their performance at same locations, we find that MZ3 significantly underperforms SGS10 in terms of 5G throughput, likely due to hardware issues of MZ3 or its 5G mod. We thus use SGS10 for all experiments. To further ensure that our experiments are not affected by device artifacts, we purchase multiple SGS10 and confirm that they exhibit similar 5G performance.
In addition, we confirm that despite 5G's high throughput, the device-side processing is not a bottleneck for SGS10, which is a high-end smartphone equipped with an \feng{octa-core CPU, 8 GB memory, Qualcomm Snapdragon 855 System-on-Chip (SoC), and X50 5G modem.}
\feng{We also use SGS10 devices to test \hd{T-Mobile} and \hd{Sprint}.
%
In addition, the SGS10 supports both 4G and 5G, allowing us to compare them on the same device.
}

\textbf{Experiment Sites.}
\feng{For most of the experiments involving \vz,}
we conduct experiments at 4 locations (\textbf{A}, \textbf{B}, \textbf{C}, and \textbf{D}).
A is a popular downtown
area in Minneapolis with many buildings.
B is at the boundary of the
5G coverage area in downtown Minneapolis.
C is inside a hotel room in downtown Chicago where we stand near an open window.
D is near the U.S. Bank stadium in Minneapolis with a large open space.
We believe that these 4 locations are representative in terms of their environment (open/crowded space, low/high surrounding buildings, indoor/outdoor, \etc).
\feng{We also conduct experiments at multiple locations in Atlanta for \vz, \hd{T-Mobile}, and \hd{Sprint}, with the details to be described in~\S\ref{sec:landscape}.}

\begin{figure*}[t]
    \centering
    \begin{minipage}{.29\textwidth}
        \includegraphics[width=1\textwidth]{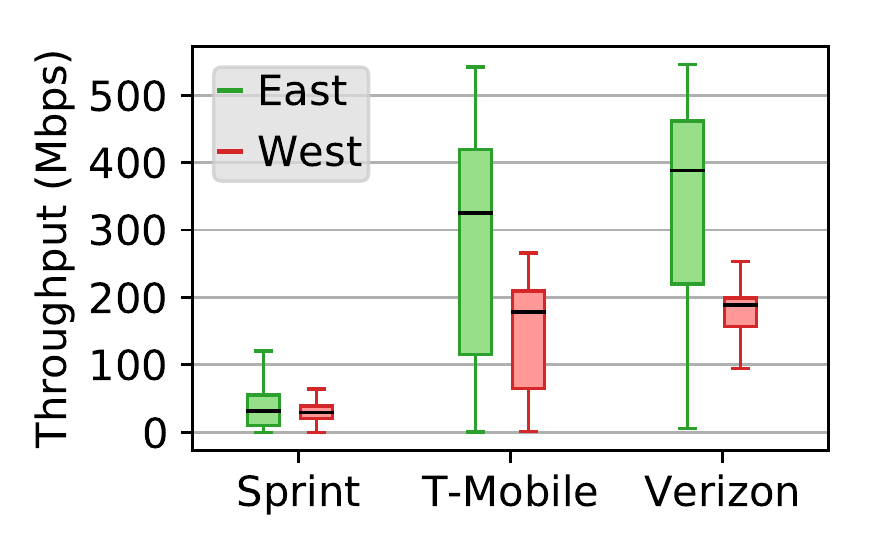}
        \vspace{-.3in}
    \end{minipage}
    \hfill
    \begin{minipage}{.29\textwidth}
        \includegraphics[width=1\textwidth]{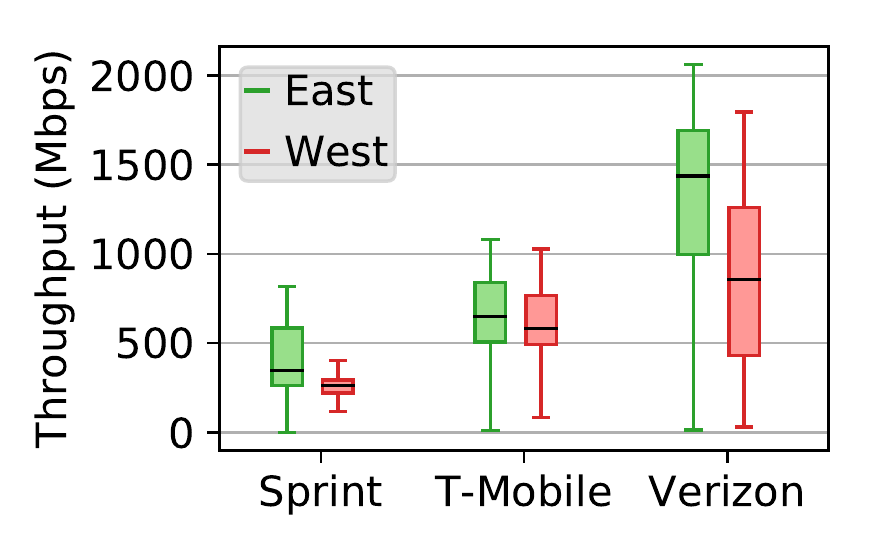}
        \vspace{-.3in}
    \end{minipage}
    \hfill
    \begin{minipage}{.29\textwidth}
        \includegraphics[width=1\textwidth]{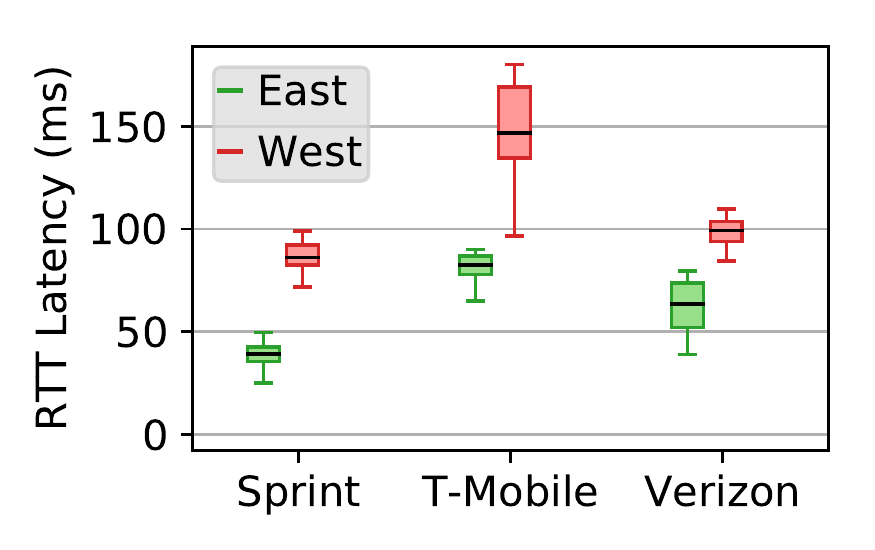}
        \vspace{-.3in}
    \end{minipage}

     \begin{minipage}{.27\textwidth}
         \caption{Throughput of a single \texttt{TCP} connection (Atlanta).}
        \label{fig:carrier_1}
        \vspace{-.1in}
    \end{minipage}
    \hfill
    \begin{minipage}{.27\textwidth}
        \caption{Aggregated throughput from 8 \texttt{TCP} connections (Atlanta).}
        \label{fig:carrier_8}
        \vspace{-.1in}
    \end{minipage}
    \hfill
    \begin{minipage}{.27\textwidth}
        \caption{Base PING RTT (Atlanta, 3 carriers).}
        \label{fig:carrier_rtt}
        \vspace{-.1in}
    \end{minipage}
\end{figure*}

\textbf{Server Selection.}
Due to the ultra-high bandwidth of 5G, the bottleneck of an end-to-end path may potentially shift from the wireless hop to the Internet -- a situation that seldom appears in 3G/4G.
Since the focus of our study is 5G, in most experiments we do \emph{not} want such a shift to occur.
\feng{To see how server selection affects the 5G performance,
we carefully experiment with several server instances offered by major cloud service providers such as Microsoft Azure, Amazon Web Service (AWS), and Google Cloud, in different locations (\eg U.S. east and west coast).
%
We observe that the server location does affect the performance. For example, compared to a west coast server, a server on the east coast typically yields higher throughput and lower latency at our test locations.
}

\feng{
For brevity, most of the experiments throughout the paper are done against a Microsoft Azure server located in the U.S. east coast.
We select this server for three reasons.
First, when downloading data from this server, we get the highest 5G throughput (statistically) compared to servers in other locations or of other cloud providers.
Second, when we perform download tests from this server to other hosts (\eg Amazon and Google cloud instances) over the wired Internet, we get $\sim$3 Gbps throughput, which is much higher than the highest 5G speed we can obtain, during different times of a day.
Third, we compare our throughput measurement results with those generated by Ookla Speedtest~\cite{ookla}, a state-of-the-art Internet speed test service, and find both results match. Note that we do not directly use Ookla Speedtest due to a lack of fine-grained test control (number of TCP connections, test duration, test automation \etc).
The above observations give us confidence that for an end-to-end path from a UE to the server, the Internet is unlikely the bottleneck.
For some experiments (\eg latency measurement and \texttt{HTTP(S)} download test), we also report other servers' results to demonstrate the impact of server selection.
%
%
%

}


\textbf{Test Workload.}
\feng{To measure transport-layer metrics including
\texttt{TCP}/\texttt{UDP} throughput, RTT, and packet loss rate,}
we perform large bulk data transfers for bandwidth probing. Specifically,
our UE issues one or more \texttt{TCP} connections or a \texttt{UDP} session to download data from an Internet server.
Since it is difficult to root our UEs, we run the cross-compiled version of \hd{iPerf} 3.6~\cite{iperf36} to measure the transport-layer metrics.
%
%
\feng{We also test important applications such as \texttt{HTTP(S)} download and web page loading over 5G,
with details to be presented in~\S\ref{sec:app_perf}.}
%

\textbf{UE-side 5G Monitoring Tool.} \arvind{At the time when we conducted this study,
the \textit{then} state-of-the-art Android OS (version 9) did not support accessing 5G-NR related information.\footnote{As of October 2019, Android 10 provides 5G-NR APIs, but no 5G phone was eligible for the update.}} We are also not aware of any dedicated public use UE-side monitoring tool for 5G networks. Due to these limitations, we develop our own tool that collects the following information to support our measurements:
(1) the UE's fine-grained location,
(2) all available network interfaces,
(3) the actively used network interface and its IP address,
(4) the cell ID (mCID) that the device is connected to,
(5) the cellular signal strength, and
(6) the 5G service status.
The above information is obtained from various Android APIs.
Regarding the last item (the 5G service status), we find that when the phone is connected to 5G,
the \hd{getDataNetworkType()} API of Android
\hd{TelephonyManager} still returns LTE.
%
From our experiments, we find the \hd{ServiceState} object when converted to a raw string representation contains the fields
\hd{nrAvailable} and \hd{nrStatus}. We find that they reliably correlate to one of the three 5G connection statuses: (1) the UE is not in a 5G coverage area (\hd{nrAvailable=F}), (2) the UE is in a 5G area but is connected to 4G due to, for example, poor 5G signals (\hd{nrAvailable=T}, \hd{nrStatus=NOT\_RESTRICTED}), and (3) the UE is connected to 5G (\hd{nrAvailable=T}, \hd{nrStatus=CONNECTED}). 
\feng{We have verified that this tool works with \vz, \hd{T-Mobile}, and \hd{Sprint}.}


\vspace{-2em}
\feng{\mysection{\hspace{-.1in} Overview of Today's 5G Performance}}
\label{sec:landscape}

We begin our study by measuring the 5G performance of commercial 5G carriers in the U.S.
%
We select a total number of \arvind{6 locations} in the downtown areas of Atlanta, where three out of the four carriers in Table~\ref{tab:carriers} offer 5G services: \hd{Verizon} (\vz), \hd{T-Mobile}, and \hd{Sprint}.
The selected locations have diverse urban environments such as high buildings, open plazas, and public transit hubs.
We perform the following experiments using SGS10: (1) a 60-second measurement of downlink throughput over a single TCP connection, (2) a 60-second measurement of downlink throughput using 8 parallel connections, and (3) \arvind{60} measurements of the base RTT (end-to-end PING without cross traffic). The methodologies are detailed in~\S\ref{sec:los}.
We experiment with two servers, one on the east coast and the other on the west coast.
\arvind{We repeat the above test 10 times for each unique (location, carrier, server) tuple. This leads to a total number of more than 350 tests.}
Figures~\ref{fig:carrier_1},~\ref{fig:carrier_8}, and~\ref{fig:carrier_rtt} plot the single \texttt{TCP} connection throughput, \texttt{TCP} throughput
aggregated from 8 connections, and the base RTT, respectively, across all the tests for the three carriers and two servers.

We make several observations.
First, in real-world urban environments, commercial 5G often exhibits great performance. For example, on \vz's network, the \texttt{TCP} downlink throughput can achieve up to 2 Gbps -- considerably better than today's top-notch residential broadband networks. On the other hand, the throughput variation is huge -- the throughput may drop to close to 0 for \vz and \hd{T-Mobile}.
Second, \hd{T-Mobile} and \vz, which use mmWave technology, provide a much higher median throughput compared to \hd{Sprint}, which operates at the mid-band frequency (2.5 GHz). This demonstrates the advantage of mmWave 5G -- its ultra high-speed (\S\ref{sec:related}).
Third, the latency offered by today's commercial 5G networks remains high -- there is little improvement over 4G. \arvind{This can be attributed to two reasons: (1) all carriers employ an NSA model that shares much 4G infrastructures with 5G, and (2) a lack of or limited use of edge support that helps shorten the end-to-end latency. We expect both to be addressed in the future in order to achieve sub-millisecond RTT.}
Fourth, we make an interesting observation that the aggregated throughput provided by 8 parallel connections is significantly higher than the throughput of a single \texttt{TCP} connection, across all three carriers. We will further investigate this in~\S\ref{sec:los}.
Fifth, the server location indeed affects both the throughput and latency (\S\ref{sec:setup}).

\arvind{We also evaluate upload performance of all the three carriers. We find that at the time of conducting this study, the upload speeds were far below the download speeds. For instance, the upload speeds of \vz and \hd{T-Mobile} peaked at $\sim$60~Mbps, while for \hd{Sprint} it was around 30~Mbps. Due to this observation (which is further confirmed in~\cite{uploadspeed}), this paper primarily studies download performance.}

Note that we do not intend to compare or rank the carriers given the limited locations we sample and given that 5G is still in its early stage. Instead, the above results provide a first impression of today's commercial 5G services, and motivate our subsequent in-depth study of 5G performance in diverse scenarios.


\feng{\mysection{5G Performance of Stationary UE}}
\label{sec:sta}

\feng{
We now closely examine how various factors such as LoS/NLoS, UE-panel distance, and mobility affect the 5G performance, through carefully designed controlled experiments. Unless otherwise noted, our experiments are conducted over \vz's 4G and mmWave 5G networks. We consider the stationary UE scenario in this section, and focus on the mobility scenario in~\S\ref{sec:mobility}.
}

\mysubsection{\texttt{TCP/UDP} Performance Under LoS}
\label{sec:los}

\begin{figure}[h]
\centering
\includegraphics[height=.17\textwidth]{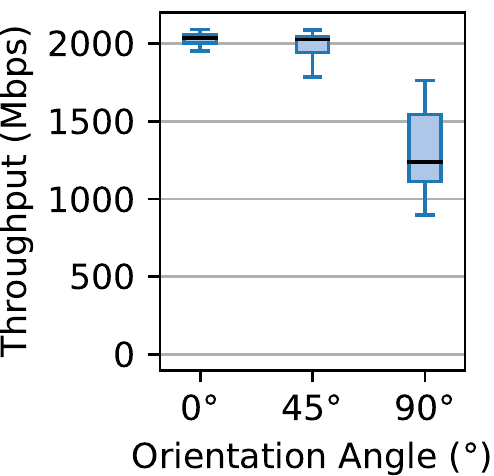}
\includegraphics[height=.17\textwidth]{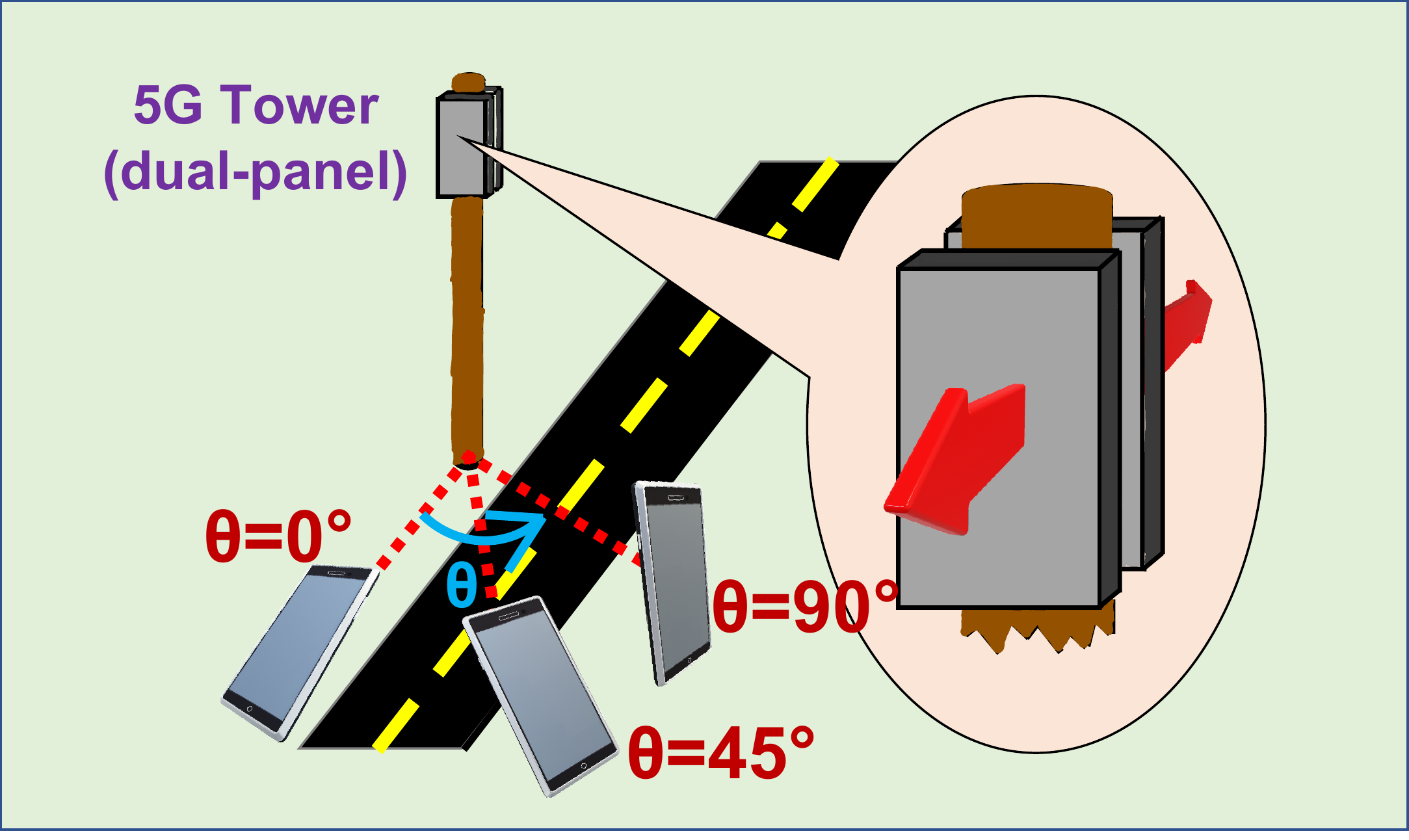}
\caption{Impact of UE-panel orientation on throughput.}
\vspace{-.15in}
\label{fig:ori}
\end{figure}

We begin with understanding 5G performance when clear LoS is present.
Specifically, we conduct experiments at
Locations A, B, and C (\S\ref{sec:setup}).
At all locations, we ensure that we can visually see the 5G panel and there is LoS between the phone and the panel. At A and B, we select 5 UE-panel distances from 13m to 75m. We use a Leica DISTO E7500i professional laser distance meter~\cite{leciatool} to accurately measure the distance. For each distance, we experiment with 3 orientations: 0\DEG, 45\DEG, and 90\DEG (see Figure~\ref{fig:ori}).
For C,
the distance (62m) and orientation (0\DEG) are fixed.

In each test, we perform \texttt{TCP} bulk download for 60 seconds using \{1, 2, 4, 8, 16\} parallel \texttt{TCP} connections, and measure the throughput and RTT every second (reported by \hd{iPerf}). For all bulk download tests, unless otherwise mentioned, we start collecting data 20 seconds after the \texttt{TCP} flow(s) start in order to mitigate the impact of \texttt{TCP} slow start.
%
We repeat the entire process 3 times. All experiments were conducted in clear weather with the phone being held in hand. We believe that the above combinations provide realistic and diverse environmental configurations of urban 5G access from smartphones. Additionally, to ensure fair comparisons
we also perform the same tests over 4G.

\begin{figure}[t]
    \centering
        \includegraphics[width=.35\textwidth]{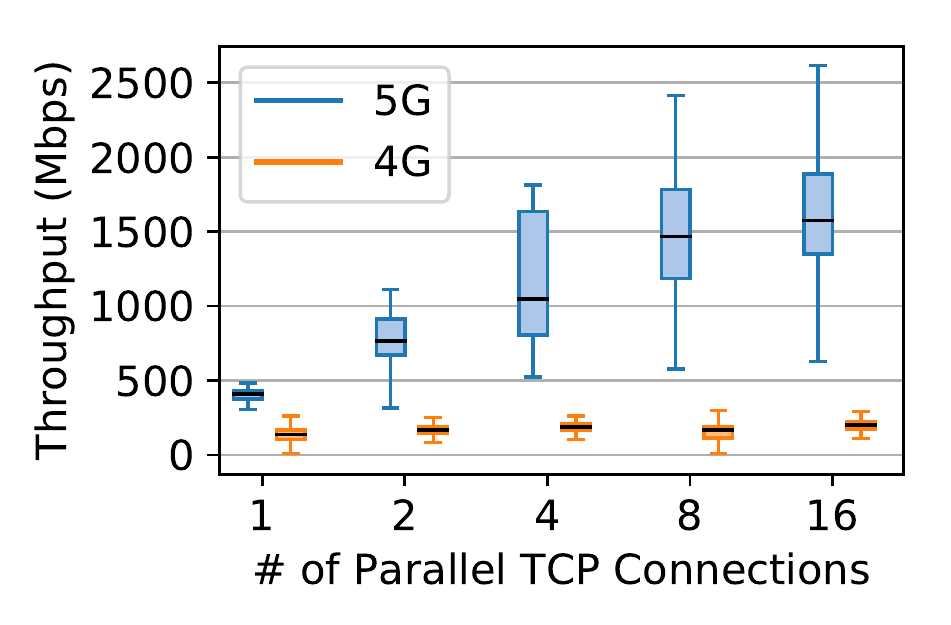}
        \vspace{-.2in}
	    \caption{\texttt{TCP} performance under LoS: throughput.}
        \label{fig:tcp_th}
        \vspace{-.15in}
\end{figure}

\begin{figure}[t]
    \centering
        \includegraphics[width=.35\textwidth]{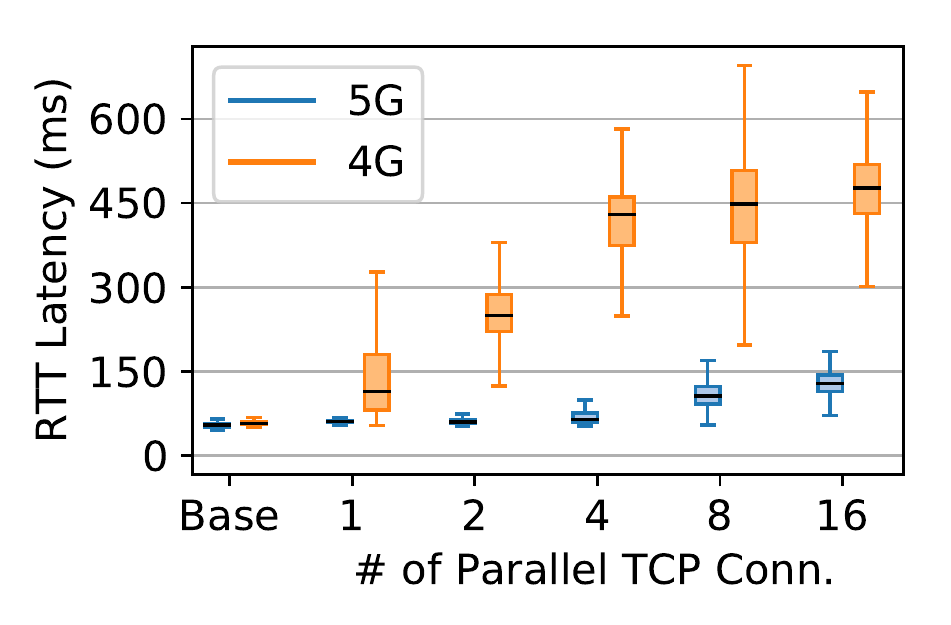}
        \vspace{-.2in}
        \caption{\texttt{TCP} performance under LoS: RTT.}
        \label{fig:tcp_rtt}
         \vspace{-.15in}
\end{figure}

Figures~\ref{fig:tcp_th} and \ref{fig:tcp_rtt} show the measurement results of throughput and RTT, respectively, for different numbers of concurrent \texttt{TCP} connections. Each box plot is across all 1-second measurement samples for a specific setup. We make several observations.
%
\arvind{First, }with 8 parallel \texttt{TCP} connections, the median 5G and 4G throughputs are \arvind{1467 and 167 Mbps}, respectively. However, 5G throughput exhibits much higher variations than 4G despite the presence of LoS.
This is due to the PHY-layer nature of 5G signals as well as potential inefficiencies at various layers.
For example, at PHY/MAC layers, smartphones' small form factor makes engineering a 5G modem challenging~\cite{mmwave_for_smartphones}.
At the transport layer, an excessive number of \texttt{TCP} connections may incur cross-connection contentions, which may also lead to throughput variation in particular for 5G whose available bandwidth is less stable compared to 4G.


Second, 5G throughput improves as the \texttt{TCP} concurrency level increases, with the bandwidth being fully utilized when there are more than 8 concurrent connections.
\feng{Recall from Figure~\ref{fig:carrier_1} and~\ref{fig:carrier_8} that such a phenomenon also occurs in other 5G carriers. Through controlled experiments, we confirm that this is 5G-specific, \ie it does not appear in wired, WiFi, or 4G networks.}
It is either because the 5G carriers are imposing per-TCP-connection rate limiting, or because they are not able to support very high throughput for a single \texttt{TCP} connection.
In either case, it may hurt the performance of single-connection protocols such as \texttt{HTTP/2}~\cite{http2_spec}.
\feng{It could also encourage application developers to aggressively increase the \texttt{TCP} concurrency that may adversely affect the \texttt{TCP} performance when the application operates in other networks.}

\begin{table}[h]
        \begin{tabular}{c|c|c|c|}
            \cline{2-4}
                        & \textbf{First Hop} & \textbf{East US} & \textbf{West US} \\
                        & \textbf{RTT (ms)}  & \textbf{Total RTT}     & \textbf{Total RTT}     \\ \hline
            \multicolumn{1}{|l|}{\textbf{5G}} & 27.4$\pm$6.4  & 54.0$\pm$4.5  & 81.9$\pm$5.5  \\ \hline
            \multicolumn{1}{|l|}{\textbf{4G}} & 29.2$\pm$4.8  & 58.0$\pm$4.3  & 88.9$\pm$5.5  \\ \hline
        \end{tabular}
\caption{\arvind{\texttt{ICMP}-based} 1st Hop RTT and impact of server's network location on total (end-to-end) PING RTT on 4G/5G. The reported numbers are averaged across all runs.}
\vspace{-.25in}
\label{tab:rtt}
\end{table}

Third, regarding the latency, 5G and 4G exhibit similar base RTT (\ie end-to-end PING) at around 56 ms for the east coast server.
\feng{As described in~\S\ref{sec:landscape}, high 5G RTTs also appear in other 5G networks, and can be explained by the NSA nature of today's 5G networks.}
To understand how much 5G contributes to this RTT,
we perform \texttt{ICMP}-based \hd{traceroute} on the UE to measure the hop-by-hop RTT.
As shown in Table~\ref{tab:rtt}, we find that the first-hop RTT, which presumably covers the
RAN (Radio Access Network), is around 28~ms for both 4G and 5G, accounting for
around 50\% of the end-to-end RTT. Changing the server location to west U.S. reduces this fraction
to around 33\%.
We then consider the RTT during a bulk transfer. 
As shown in Figure~\ref{fig:tcp_rtt},
4G RTT inflates drastically because of its deep in-network buffers~\cite{jiang2012tackling,huang2013depth}. Bufferbloat in 5G is much less severe, likely due to the fast 5G speed that drains the buffer much faster than 4G.
Also \vz 5G exhibits low packet loss rates, with the 50\%, 75\%, and 99\% percentiles being 0.01\%, 0.1\%, and 1.2\%, respectively.

\textbf{Concurrent Clients.}
We also find that the bandwidth available from a single 5G panel is shared among its associated devices. For a simple test, we place two SGS10 side-by-side (ensuring both are connected to the same panel with LoS).
We first let only one device run \hd{iPerf} with 8 parallel connections. After \texttt{TCP} slow start, the maximum throughput is around 1.8 Gbps. We then let both perform \hd{iPerf} simultaneously, and find that
the above maximum throughput is approximately evenly divided between both devices
(avg.~$\approx$882~Mbps for one and $\approx$931~Mbps for the other).
We consistently observe this in several locations at \arvind{different} times.
This finding suggests that as the 5G user base increases, the perceived throughput may drop significantly due to bandwidth sharing among concurrent clients. However, due to very limited 5G users today, it is hard to quantify the impact. For most if not all experiments reported in this study, we have high confidence that our phone is the sole device connected to the 5G NR. In this sense, our study
provides an important ``baseline'' for studying how 5G performance evolves in the future as the user base increases.




\textbf{\texttt{UDP} Performance Under LoS.}
We repeat the measurement in~\S\ref{sec:los} using \texttt{UDP}. Since \texttt{UDP} does not provide congestion control, we manually increase the sending rate exponentially from 512 Kbps to 2 Gbps. we find that for sending rates up to 1 Gbps, the receiver-side loss rate is close to 0.
This indicates \vz 5G's compatibility with \texttt{UDP}-based protocols such as \texttt{QUIC}~\cite{quic} and \texttt{HTTP/3}~\cite{http3_draft} at these low to medium data rates.
However, at our test locations, \vz's 5G is not able to reliably sustain 1 Gbps or a higher sending rate over \texttt{UDP}, as we observe a packet loss rate of up to 17\% at 1 Gbps.

\vspace{-1em}
\mysubsection{Impact of the Environment}
\label{sec:nlos}

\begin{figure}[h]
\centering
\includegraphics[width=0.47\textwidth]{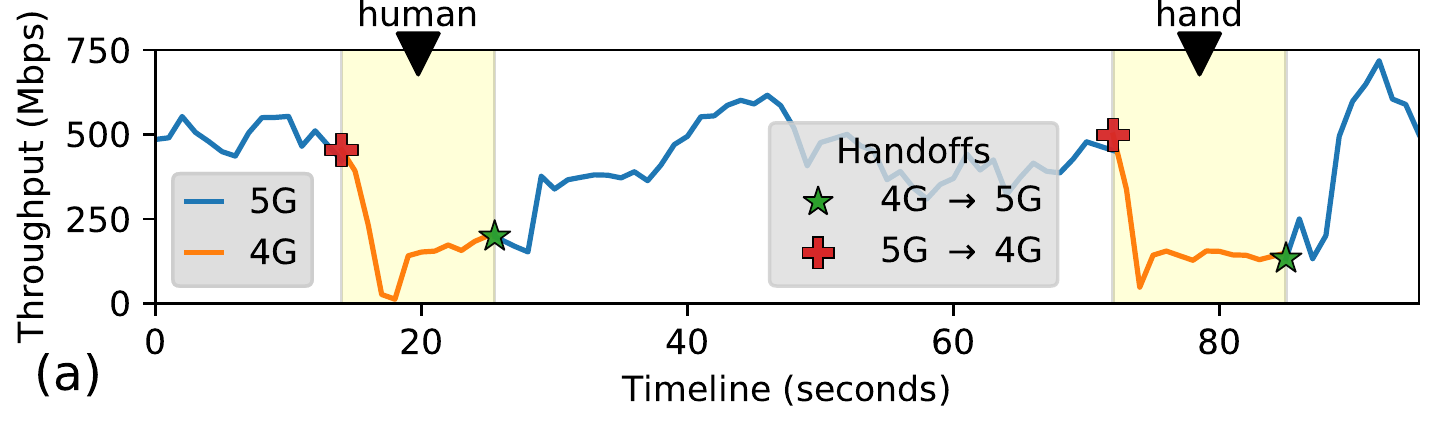}
    	\label{fig:obstruction-indoor}
\includegraphics[width=0.47\textwidth]{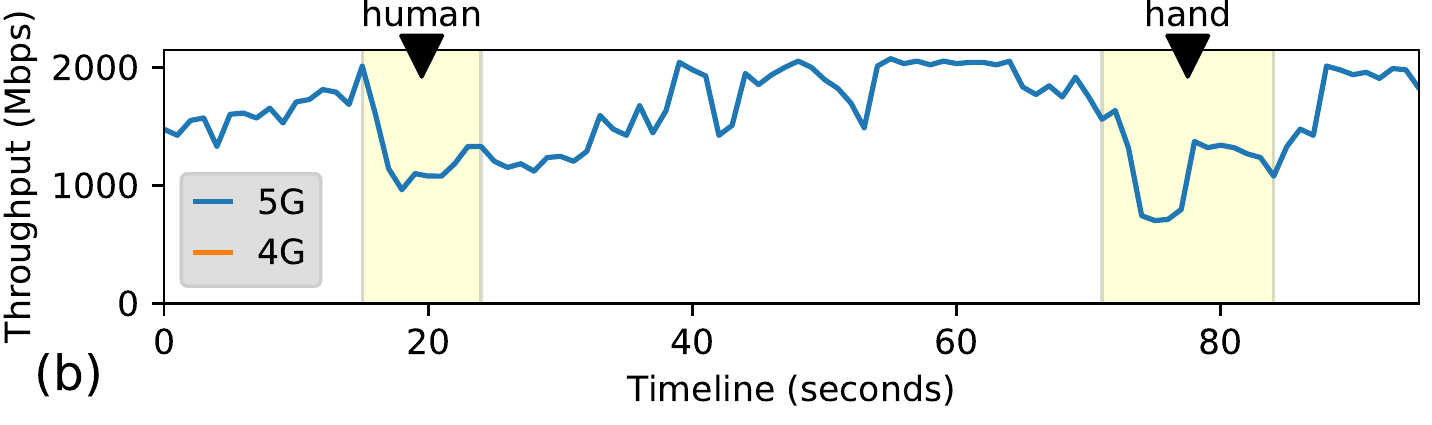}
    	\label{fig:obstruction-outdoor}
    \vspace{-.2in}
\caption{Obstruction tests. (a) Location C with ineffective multipath, (b) Location A with effective multipath.}
\vspace{-.1in}
\label{fig:obs}
\end{figure}

\textbf{Obstruction and NLoS Performance.}
The experiments in~\S\ref{sec:los} assume a clear LoS path without any obstruction.
We now place different types of objects along the LoS path to test whether 5G signals can penetrate/bypass them.
%
We first examine two common obstructions: human body and hand.
\arvind{We stay at Location C (the Chicago hotel room) with an awning window directly facing a 5G-NR panel (62m). When the window is opened, the phone has LoS to the 5G-NR panel.} We start a bulk transfer over 5G with 8 parallel \texttt{TCP} connections. During the data transfer, we block the LoS path with a human body and then a hand. We repeat this experiment 10 times and observe qualitatively similar results, with one representative run illustrated in Figure~\ref{fig:obs}(a). As shown, both obstructions trigger 5G-to-4G handoffs (\S\ref{sec:hand}) and lead to significant performance degradation.
In contrast, when experimenting with 4G, neither blockage incurs noticeable throughput drop (figure not shown), due to the low frequency of 4G signals.

The above results show that it is difficult for 5G signals to penetrate a hand or human body, causing NLoS between the transmitter and receiver.
We then study other types of obstructions using similar methods. We find that when the UE is inside a backpack, a cardboard box, or a clear glass, 5G signals can penetrate these containers (experimented with $<$100 meters distance to the 5G panel with LoS). However, 5G signals can hardly penetrate human bodies, trains, pillar structures, and tinted glass.
We find that 5G works in vehicles since the front windshield is typically clear glass. 


We repeat the same experiment in Location A,
also using a human body and a hand as obstructions. As shown in Figure~\ref{fig:obs}(b), the impact of the obstructions becomes smaller: the 5G connectivity persists despite a fair amount of performance degradation.
Figures~\ref{fig:obs}(a) and~\ref{fig:obs}(b) indicate that
the environment can affect the impact of obstructions.
At Location A, despite the NLoS created by the obstructions,
the nearby buildings
can reflect signals and create multiple wireless paths, and the reflected signals can still reach the UE. 
At Location C, the room has UV-protective windows that are very common in today's buildings.
Since the windows attenuate reflected 5G signals~\cite{zhao201328}, multipath becomes ineffective.
In other words, the only effective signal propagation path is through the open window. Blocking it inevitably degrades the performance. Note that during the experiment we only open the awning window and keep other windows closed.

\textbf{Impact of UE-Panel Orientation.}
We also investigate how the UE's orientation to the panel affects network performance.
We define the orientation as the minimum angle between the LoS and all normal vectors of the base station's panels. As illustrated in Figure~\ref{fig:ori}, an orientation of 0\DEG is preferred because the panel is directly facing the UE, while an orientation of 90\DEG is the least ideal case.
Our orientation test is performed at Location D where we can find a large LoS area centered by a 5G tower. We pick three spots whose orientations are 0\DEG, 45\DEG, and 90\DEG. All spots have a 25m distance to the tower.
At each spot, we perform three 60-second bulk download tests using 8 parallel connections.
As shown in Figure~\ref{fig:ori}, we observe very small performance difference between 0\DEG and 45\DEG orientation, attributed to
the environmental reflection and beamforming.
However, in the extreme case where the orientation becomes 90\DEG, we do observe a median throughput drop of 40\%.


\textbf{Impact of UE-Panel Distance.}
We study how the distance between a UE and the panel affects network performance. We conduct the experiment at Location B, where we select five spots with their distances to the panel being 25m, 50m, 75m, 100m, and 160m, respectively. The panel and the five spots are on the same line. At each spot, we conduct three 60-second bulk download tests using 1 and 8 parallel \texttt{TCP} connection(s). During all tests, we ensure that the UE is associated with the same panel, \ie there is no 5G-4G or 5G-5G handoff (\S\ref{sec:hand}).
%
We find that at Location D (near the stadium), a UE can also reliably connect to the same panel at a long distance. So we conduct tests there as well using a similar setup.

\begin{figure}[t]
    \centering
    \includegraphics[width=.65\linewidth]{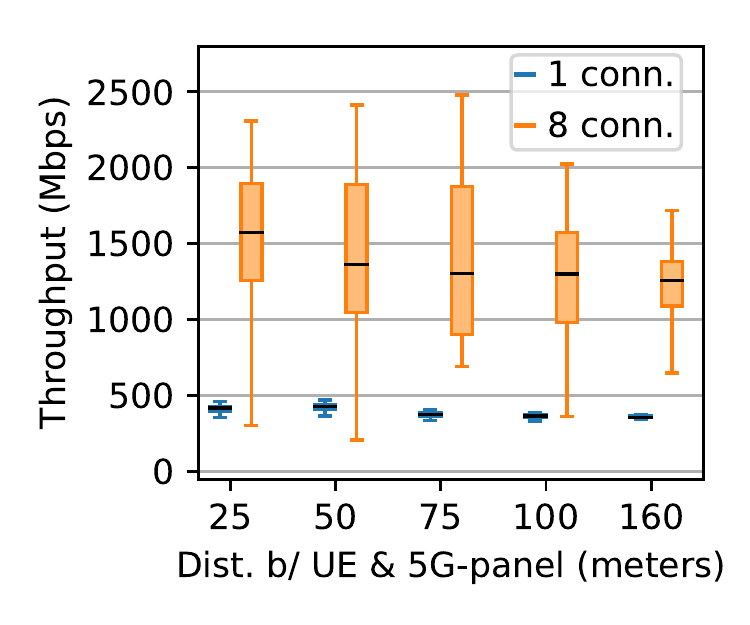}
    \vspace{-.2in}
    \caption{Impact of UE-Panel distance on throughput.}
   \vspace{-.2in}
    \label{fig:dist-all}
\end{figure}

Figure~\ref{fig:dist-all} plots the throughput distributions of different distances, where each box is across all 1-second samples measured at Locations B and D.
As shown, the throughput moderately reduces as the distance increases. For 8 parallel \texttt{TCP} connections, the median throughput decreases by 17\% (20\%) at 100m (160m) compared to that at 25m. We attribute this moderate reduction to
the clear LoS and sufficiently high transmission power of 5G antennas. For a single \texttt{TCP} connection, no noticeable throughput drop is observed because of bandwidth under-utilization.


\textbf{Impact of Rain.}
Common weather conditions such as rain and snow also affect mmWave performance as its signals are weakened by particles or even moisture~\cite{design_5g_beamforming_antennas} in the air.
\arvind{During days with moderate rainfall (between 1.0mm to 2.8mm per-hour)\cite{weather1,weather2},} we conduct 2-hour measurements at Location B and near Location C
(25m and 50m distance, 0\DEG orientation, 8 parallel \texttt{TCP} connections). We then compare the \texttt{TCP} throughput with that measured on sunny days at the same location with the same setup. We find that at 25m, the median throughput during the rain drops by only 3\%, while at 50m, the median throughput reduction is around 30\%. The results confirm that rain does degrade 5G throughput.

Overall, among all factors (obstruction, orientation, distance, and weather),
obstruction typically
incurs the highest impact on mmWave 5G performance.
Fortunately, in urban environments, surrounding signal reflections can oftentimes mitigate the performance degradation, allowing 5G to function under NLoS.
\feng{Based on our experiments, we find that our findings in this subsection also qualitatively apply to \hd{T-Mobile}'s 5G networks. In contrast, for \hd{Sprint}, the environment incurs a much smaller impact on the performance due to the omni-directional nature of its radio signal.}



\feng{\mysection{Mobility and Location-awareness}}
\label{sec:mobility}

\feng{Mobility poses a major challenge for mmWave: it will cause continuous changes of the UE-panel distance and orientation; it may incur random LoS/NLoS switches due to obstacles; in particular, it will also lead to more frequent handoffs compared to those in 4G. In this section, we first characterize handoff and mobility performance for 5G (\S\ref{sec:hand}). We then investigate the feasibility of using a UE's fine-grained location to predict mmWave 5G performance (\S\ref{sec:loc}).
}

\feng{\mysubsection{Handoff and Mobility Performance}}
\label{sec:hand}

Handoffs in 5G differ from those in 4G/3G in both the horizontal and vertical dimensions.
A Horizontal Handoff (HH) occurs when a UE's association switches from one panel (in 5G's term) to another. In 5G, HHs may frequently occur due to the smaller coverage of 5G panels compared to 4G towers.
A Vertical Handoff (VH) is triggered when the wireless technology changes (\eg 5G to 4G). VHs are also prevalent in 5G
whose signals are more unstable than 4G.
%

We closely examine \vz's handoff mechanisms. In 5G NSA, a UE may be in one of the three states:
(1) the UE is connected to 5G,
(2) the UE is in a 5G coverage area but is connected to 4G due to, for example, poor 5G signals, and
(3) the UE is not in a 5G area.
Recall that~\S\ref{sec:setup} details how they are identified by our monitoring tool.
We refer to these states as
\textbf{C} (Connected to 5G), \textbf{R} (Ready for 5G but not yet connected), and \textbf{O} (Outside 5G coverage), respectively.
We use this state and the cell ID (also collected by our tool) to track both HH and VH.
Note that in 5G, cell IDs identify 5G-NR panels.
%
%

\begin{table}[t]
\begin{tabular}{|c|c|c|}
\hline
\textbf{Type} &    \textbf{Description}                      & \textbf{Sequence} \\ \hline
$P1$   & VH, 5G$\rightarrow$4G, same cellID   & C1$\rightarrow$R1     \\
$P2$   & VH, 4G$\rightarrow$5G, same cellID   & R1$\rightarrow$C1     \\
$P3$   & VH+HH, 5G$\rightarrow$4G, diff cellID   & C1$\rightarrow$R2     \\
$P4$   & HH, 4G$\rightarrow$4G, diff cellID   & R1$\rightarrow$R2     \\
\hline \hline
$O1$   & \multicolumn{1}{c|}{5G temporarily disrupted, same tower}  & P1 $\rightarrow$ P2 \\ 
$O2$   & \multicolumn{1}{c|}{5G to 5G between two panels}           & P3 $\rightarrow$ P2 \\ 
$O3$   & \multicolumn{1}{c|}{5G to 5G between two panels}           & P1 $\rightarrow$ P4 $\rightarrow$ P2 \\
$O4$   & \multicolumn{1}{c|}{5G to 4G between two panels}           & P1 $\rightarrow$ P4 \\ 
$O5$   & \multicolumn{1}{c|}{4G to 5G between two panels}           & P4 $\rightarrow$ P2 \\ 
\hline
\end{tabular}%
\caption{Primitive (top) \& combinational (bottom) handoffs.}
\vspace{-.35in}
\label{tab:handover}
\end{table}

We then conduct experiments in the three cities under various mobility levels (stationary, walking, and driving) to capture the above data related to handoffs. We identify 4 types of \emph{primitive handoffs} ($P1$ to $P4$) as listed in the upper part of Table~\ref{tab:handover}.
A primitive handoff is between the C and R states as described above.
%
$P1$ and $P2$ are VHs because they are handoffs between 4G and 5G.
%
When a UE's 5G signal strength drops (\eg due to a NLoS obstruction), $P1$ is triggered to downgrade the connectivity from 5G to 4G; when the network condition improves, the connectivity will be restored back to 5G ($P2$).
Note that in the 5G-ready (R) mode, the UE actually connects to a 4G radio that is on the same tower (where the 5G-NR panel resides) or a nearby tower, but the cell ID does not change.
This is likely because of NSA where 4G and 5G are deeply coupled.
At the R state, the UE is still closely monitoring the original 5G panel for a possible 4G to 5G upgrade.
%
$P3$ is similar to $P1$ except that the 5G to 4G downgrade ends at a different cell ID (panel).
$P4$ is a 4G to 4G HH from one cell ID to another.
We do not observe a C1$\rightarrow$C2 or R1$\rightarrow$C2 sequence in our data.
This is because NSA uses 4G for control-plane signaling -- the UE will need to first associate with the new cell's 4G radio for control message exchanges before establishing the 5G data channel.

Interestingly, we find that the (in)activity of user traffic can also trigger 4G-5G handoffs. A $P1$ handoff will occur when there is an inactivity of user traffic for $\sim$10 seconds; at the R state, any user traffic will restore the 5G connectivity through a $P2$ handoff, if the 5G signal is good. The rationale of such traffic-guided handoffs is to reduce the 5G standby time that may consume additional energy.

From our data, we observe that oftentimes the primitive handoffs form complex sequences that we call \emph{combinational handoff sequences}. We identify them by clustering primitive handoffs using an interval threshold (set to 10 seconds).
They are exemplified at the bottom part of Table~\ref{tab:handover} as $O1$ to $O5$.
These combinational sequences correspond to
high-level events that cannot be realized by a single primitive handoff.
For example, $O3$ represents a 5G-to-5G handoff that consists of three primitive handoffs: a 5G to 4G VH on the old panel, a 4G to 4G HH from the old to new panel, and a 4G to 5G VH on the new panel. The whole procedure takes several seconds to finish. \feng{We find that \hd{T-Mobile} employs similar mechanisms for 4G/5G and 5G/5G handoffs.}


\begin{figure}[h]
  \centering
    \includegraphics[width=.47\textwidth]{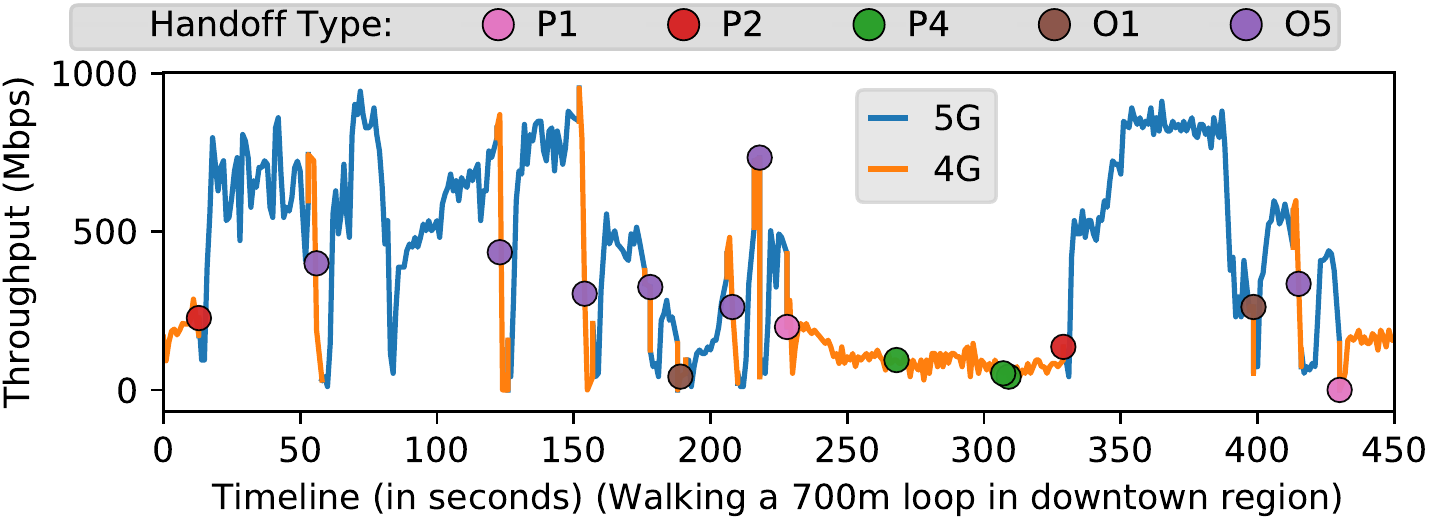}
    \vspace{-.15in}
  \caption{5G throughput and handoffs under low mobility.}
  \vspace{-.1in}
   \label{fig:walk}
\end{figure}

\feng{\textbf{Low Mobility Performance (Walking).}}
We next show a case study to demonstrate the impact of handoffs. In this experiment, one of the authors holds a phone while walking at a normal speed ($\sim$5 km/h) at Location A for about 8 minutes. The phone keeps downloading data from a server over 8 parallel connections. Figure~\ref{fig:walk} plots the throughput, cellular connectivity (4G/5G), and handoffs. During this 8-minute walk, the phone experiences 31 primitive handoffs and bounces between 4G and 5G for 13 times. Such frequent switches make the throughput highly fluctuating, ranging from 0 to 954 Mbps.
This may confuse applications (\eg video rate adaptation logic~\cite{jiang2012improving,yin2015control,mao2017neural}) and bring highly inconsistent user experiences.
These results highlight the need for cross-layer efforts that improve 5G performance under (even low) mobility.
They include PHY/MAC enhancements for reducing the handoff frequency, and
robust upper-layer solutions that can adapt to frequent 4G/5G handoffs, such as MPTCP~\cite{nikravesh2016depth} and prefetching~\cite{higgins2012informed}.

\begin{figure}[t]
    \centering
    \includegraphics[width=0.45\textwidth]{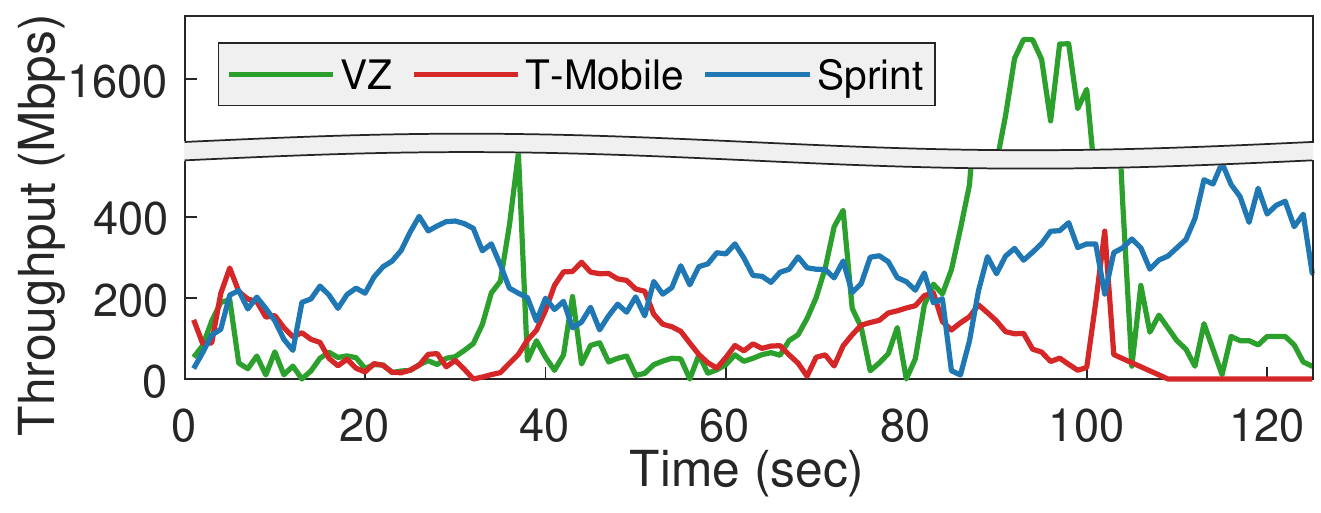}
    \vspace{-.18in}
    \caption{Driving test performance across 3 carriers.}
    \vspace{-.15in}
    \label{fig:carrier-driving}
    \vspace{-.5em}
\end{figure}

\feng{\textbf{Medium Mobility Performance (Driving).}
Besides walking, we also conduct experiments to study the 5G performance under medium mobility. We put three SGS10 devices near the front windshield of a vehicle, and then drive in midtown Atlanta while measuring the \hd{iPerf} performance (using 8 TCP connections) for \vz, \hd{T-Mobile}, and \hd{Sprint} from the three devices. The driving speed was between 20 to 50 km/h. Figure~\ref{fig:carrier-driving} shows representative 2-minute throughput traces simultaneously measured from the three carriers. As shown, both \vz and \hd{T-Mobile} exhibit significant throughput fluctuation, with the throughput often dropping close to 0, due to frequent handoffs and blockages incurred by nearby buildings and vehicles. The glasses and body of our car also weaken the mmWave signal.
We also find that \hd{T-Mobile} handles handoffs more poorly than \vz (\eg at $t$=110 s, a handoff causes the TCP connections to completely disconnect).
In contrast, \hd{Sprint} exhibits higher throughput and lower throughput variation during most of the time, due to the omni-directional nature of its mid-band frequency. 

\textbf{Multipath over 4G and 5G.}
Given the stability of omni-\linebreak directional radio in mobility compared to mmWave, a possible way to improve mmWave 5G mobility performance is to use mmWave 5G and 4G simultaneously (\eg through  MPTCP~\cite{nikravesh2016depth}) where 4G can help guarantee the basic data connectivity. To illustrate this idea, we conduct the following experiment.
We walk around the Commons Park (a 4.2-acre park in downtown Minneapolis) loop twice per day at different times of a day, for 30 days. During a walk, we capture 5G (including its 4G fallback) and 4G \hd{iPerf} throughput from our phones that are held naturally by a user. Meanwhile, the phones also log the location (using GPS)
and handoff events. A typical walk takes about 11 to 14 minutes to complete.
We then group the throughput samples into 2-meter segments (bins) along the loop. Figure~\ref{fig:variance} plots the average and standard deviation (the shaded region) of the 4G/5G throughput samples in each 2-meter segment. As shown, 4G performance is indeed more stable than 5G during an average walk:
\hd{StdDev}/\hd{Mean} across per-segment average throughput is 0.42 and 0.57, respectively, for 4G and 5G.
In addition, at certain locations such as those around the 50-th and 75-th segment, 4G outperforms 5G when 5G experiences a blackout. We expect that multipath solutions such as MPTCP over 5G and 4G can mitigate the blackout impact by seamlessly diverting the application traffic from 5G to 4G during a 5G blackout. 

}

\feng{\mysubsection{Location-based Performance Estimation}}
\label{sec:loc}

\feng{
In the 3G/4G era, location is known to be useful for predicting the cellular performance~\cite{schulman2010bartendr,margolies2016exploiting}.
%
The prediction, if reasonably accurate, can be utilized in many ways to boost the performance and resource efficiency. We now investigate location-based performance prediction in the mmWave 5G context. The rationale is that fine-grained UE location, which can be easily obtained (at least outdoor through GPS), encodes rich information about the surrounding environment influencing the network performance.

We use the Commons Park walk loop data (Figure~\ref{fig:variance},~\S\ref{sec:hand}) to investigate the feasibility of location-based performance estimation for mmWave 5G.
We find that at a given location, 5G exhibits a statistically higher throughput variation compared to 4G. To show that, let \hd{StdDev}$(s)$ and \hd{Mean}$(s)$ be the stddev and mean of all the throughput samples belonging to a given location $s$ (a 2-meter segment); we compute the average of \hd{StdDev}$(s)$/\hd{Mean}$(s)$ across all segments to be 0.70 for 4G and 1.07 for 5G.

\begin{figure}[t]
    \centering
    \includegraphics[width=0.39\textwidth]{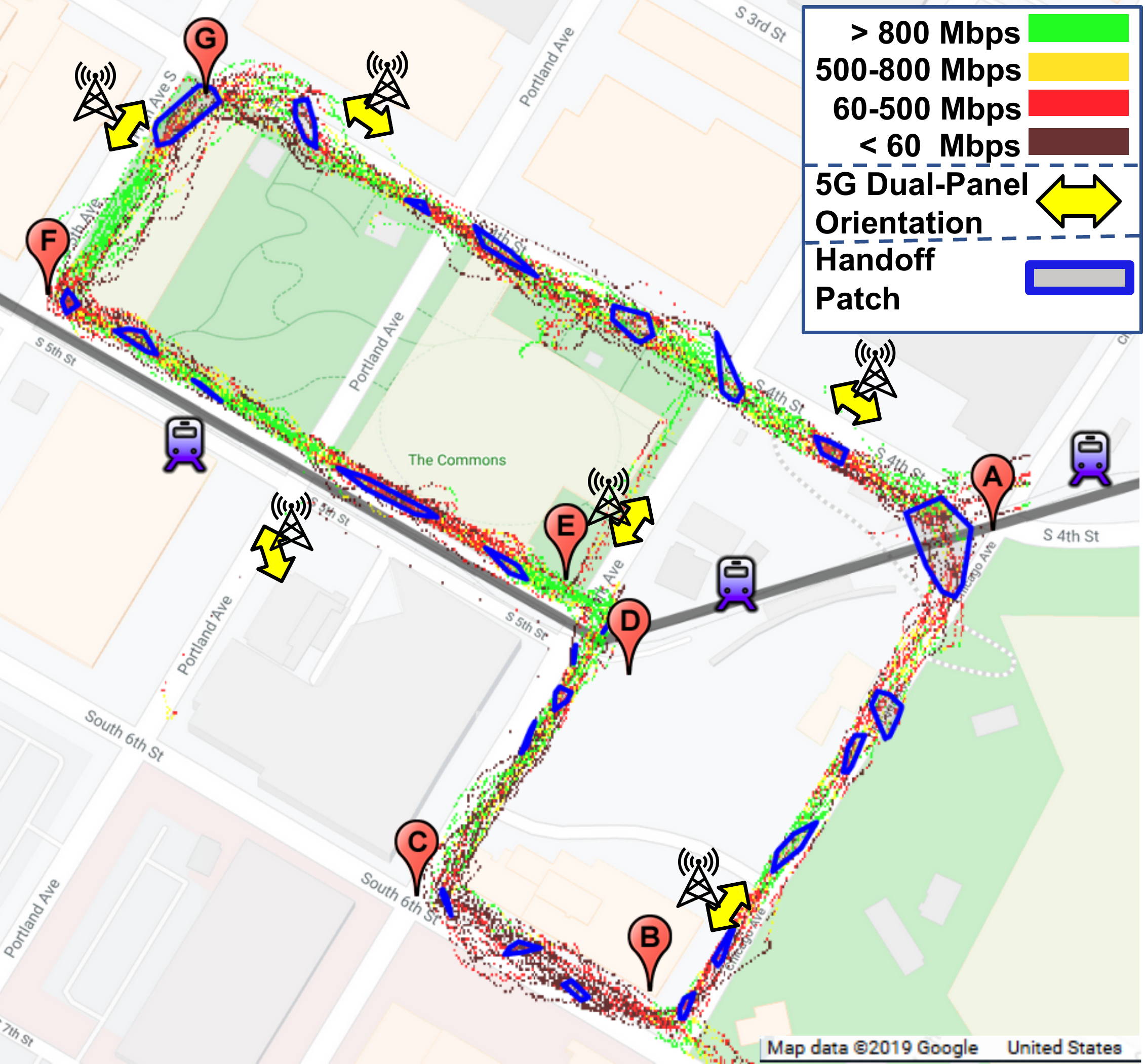}
    \vspace{-.1in}
    \caption{Throughput heatmap and handoff regions.}
    \label{fig:heat}
\end{figure}

\begin{figure}[t]
    \centering
    \includegraphics[width=0.43\textwidth]{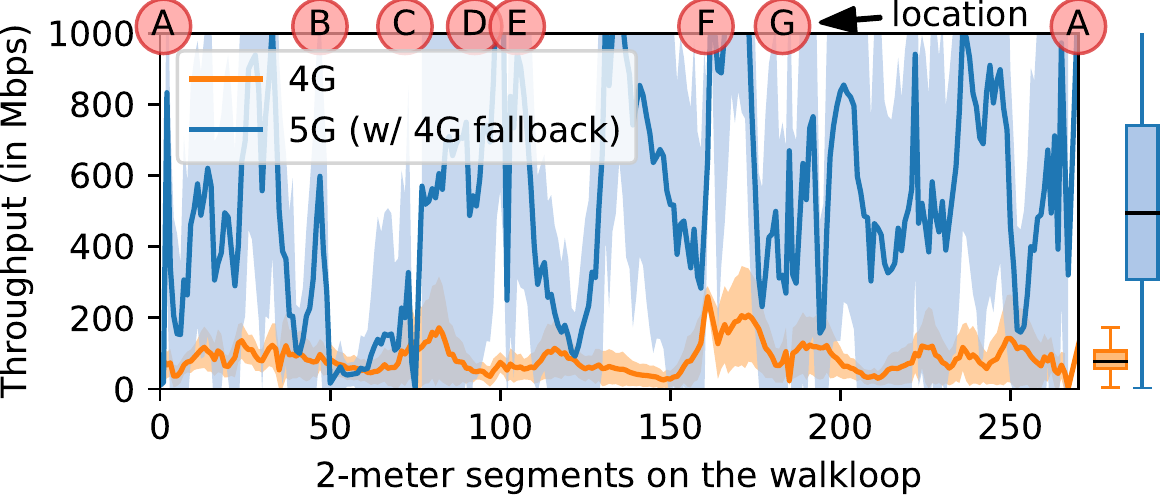}
    \vspace{-.1in}
    \caption{4G/5G throughput statistics measured from a 30-day field study in the Commons Park. The A to G labels match the corresponding ones in Figure~\ref{fig:heat}.}
    \vspace{-.1in}
    \label{fig:variance}
\end{figure}

\begin{figure*}[t]
    \centering

    \begin{minipage}{.6\textwidth}
    \includegraphics[width=1\textwidth]{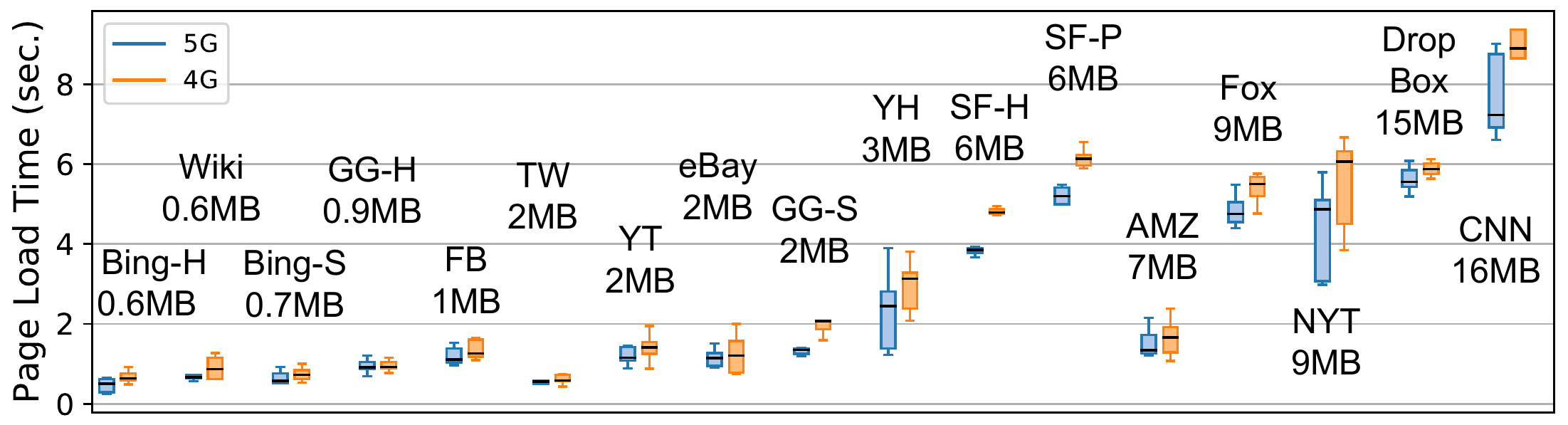}
    \vspace{-.3in}
    \end{minipage}
    \hfill
    \begin{minipage}{.28\textwidth}
    	\centering
    	\includegraphics[width=1\textwidth]{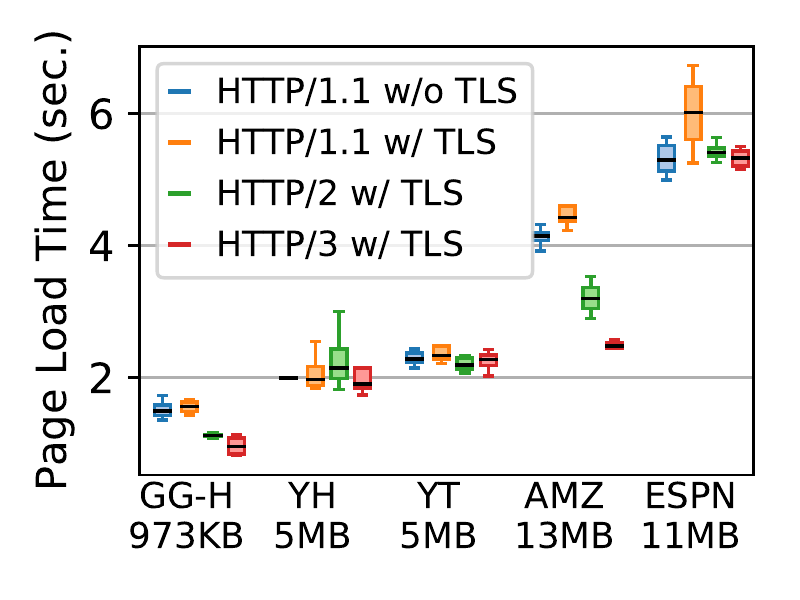}
        \vspace{-.3in}
    \end{minipage}

    \begin{minipage}{.63\textwidth}
      	\caption{4G/5G PLT over 17 pages: Bing Home, Wikipedia, Bing Search, Google Home, Facebook, Twitter, YouTube, eBay, Google Search, Yahoo, Salesforce Home, Salesforce Product, Amazon, FoxNews, NYTimes, DropBox, CNN.}
   		\vspace{-.1in}
    	\label{fig:pageload}
    \end{minipage}
    \hfill
     \begin{minipage}{.33\textwidth}
    	\caption{5G PLT for different \texttt{HTTP} versions across 5 landing pages: Google, Yahoo, YouTube, Amazon, and ESPN.}
    	\vspace{-.1in}
    \label{fig:plt_5g_controlled}
    \end{minipage}
\end{figure*}

In Figure~\ref{fig:heat}, we further visualize the 5G throughput samples as follows.
We divide the area into 1m$\times$1m grids, and distribute the throughput samples into the grids based on the location. For each grid, we compute the average throughput, based on which we assign one of the four throughput levels to the grid. We then visualize each grid using a color associated with its throughput level. The blue patches denote the handoff regions that are generated by performing DBSCAN clustering~\cite{ester1996density} for all locations where handoffs occur, and constructing the convex hull for each cluster.
%
Figure~\ref{fig:heat} exemplifies two sides: (1) there are some locations where the throughput levels appear to be homogeneous, such as the ``dead zone'' between B and C due to poor 5G coverage, the good performance zone above F due to
simple flat structures facilitating beamforming/LoS, and the bad performance zone at street corner F/G likely due to the complex environment;
(2) on the other side, in many other locations there is no clear pattern, such as many spots on the street between A and G and those between C and D.

Our finding suggests that location-based throughput estimation in mmWave 5G faces challenges due to
mmWave's sensitivity to the environment. A small perturbation
(\eg device orientation, nearby moving objects, humidity, and even the phone case~\cite{phone_case_5g})
can affect the performance, making the prediction difficult.

\textbf{Towards Context-aware Performance Prediction.}
Given the inefficiency of location-based prediction, it can be further generalized into \emph{context-aware} prediction, where a data-driven model uses a wide range of contextual and environmental data such as location, time, mobility level, orientation, weather, and traffic information to model and predict mmWave 5G performance that is inherently sensitive to the environment.
For example, we find that around Location A in Figure~\ref{fig:heat}, a passing train can cause a sudden loss of LoS to the tower. The train schedule information can thus be utilized to facilitate the context-awareness.
%
Data collected by the client can be synthesized at the base station to train a model, which will then be leveraged for robust performance prediction.
This approach can potentially outperform the location-only approach and become an important building block that guides decision making at various layers, such as
adaptive beam forming, preemptive handoff, multipath scheduling, and smart prefetching.
Meanwhile, there also exist challenges for realizing such a framework, including
potentially sparse data, data collection overhead, and privacy issues.
}

\vspace{-.5em}
\mysection{Application Performance}
\label{sec:app_perf}

\feng{This section characterizes application performance over 5G. We investigate two important applications:  web page loading and \texttt{HTTP(S)} download.
}

\vspace{-.5em}
\mysubsection{Web Page Loading}
\label{sec:web}

The World Wide Web is a critical piece in the mobile ecosystem.
We utilize the mobile Chrome browser to load the \feng{landing or content pages of 17 popular websites} listed in Figure~\ref{fig:pageload} over 5G and 4G. We then compare their page loading time (PLT).
We conduct the experiments at Locations A and C with SGS10 over 5G and 4G.
In Location A (C),
the UE-panel distance is 50m (62m) with LoS.
We connect the phone to a laptop and use the Chrome remote debugger~\cite{chrome_remote_debugger} to automate our tests and to clear the browser's cache before each loading. We load each page back-to-back over 5G and 4G, and repeat this 4 times for all sites per day, over a few weeks.

%
As shown in Figure~\ref{fig:pageload}, for most sites with small page sizes ($\le$ 3 MB), 4G and 5G achieve similar PLT. This is attributed to three reasons.
First, web browsing requires a synergy between network transfer and local processing, with the latter often being the bottleneck for small web pages particularly~\cite{wang2013demystifying}.
%
%
\feng{Second, compared to 4G, today's 5G does not bring improvement on RTT, which is known to be important for web browsing~\cite{agababov2015flywheel}.}
Third, as we load the pages from the original content providers, the bottleneck may shift from the wireless hop to the Internet.
\feng{For large pages ($>$3 MB), loading them over 5G does shorten the median PLT by 5.5\% to 19.8\% (median: 18.7\%), because of the reduction of the content fetch time.}

\feng{

We next investigate the impact of different web protocols. We capture the landing pages of five websites and host them on our controlled server (an Azure server in the east coast running OpenLiteSpeed~\cite{openlitespeed}). We then fetch the pages over 5G with four configurations: \texttt{HTTP/1.1} without \texttt{TLS}, \texttt{HTTP/1.1} with \texttt{TLS}, \texttt{HTTP/2} with \texttt{TLS}, and \texttt{HTTP/3} (\texttt{QUIC}) with \texttt{TLS}.
The experimental locations (A and C) are the same as those for Figure~\ref{fig:pageload}.
At each location, for each website, we perform 120 back-to-back page loads  over 5G in a random order across the four \texttt{HTTP} versions (30 for each version).

As shown in Figure~\ref{fig:plt_5g_controlled}, the four \texttt{HTTP} versions indeed exhibit different performance.
Recall that from the networking perspective, the major differences among \texttt{HTTP/1.1}, \texttt{HTTP/2}, and \texttt{HTTP/3} are the transport layer protocol \arvind{(\texttt{HTTP/1.1} and \texttt{HTTP/2} use \texttt{TCP} while \texttt{HTTP/3} uses \texttt{UDP})} and the connection management scheme (\texttt{HTTP/1.1} uses concurrent \texttt{TCP} connections while \texttt{HTTP/2} and \texttt{HTTP/3} employ a single connection per domain).
For four out of the five websites, \texttt{HTTP/2} outperforms \texttt{HTTP/1.1} (with \texttt{TLS}) mostly due to \texttt{HTTP/2}'s multiplexing feature that consolidates traffic onto one \texttt{TCP} connection per domain, leading to better bandwidth utilization~\cite{qian2015tm}.
Recall from~\S\ref{sec:los} that a practical issue in commercial 5G is its poor single \texttt{TCP} connection throughput compared to using many concurrently.
We find this limitation does not affect \texttt{HTTP/2} and \texttt{HTTP/3} performance in our experiments
because the small data size, the \texttt{TCP} slow start, and the computation overhead often make the bandwidth under-utilized even over a single connection.
\texttt{HTTP/3} outperforms \texttt{HTTP/2} for 2 websites and achieves similar median PLT for the other 3 websites. This is attributed to its reduced handshake time as well as its usage of \texttt{UDP} that eliminates the receiver-side HoL blocking across streams~\cite{langley2017quic}.

Overall, we find that compared to 4G, today's 5G brings benefits only for large sites with rich content; meanwhile, the optimizations brought by \texttt{HTTP/2} and \texttt{HTTP/3} are effective over 5G.
Based on our findings, we identify key improvements for reducing the PLT over 5G: (1) upgrading from NSA 5G to SA 5G to reduce the network latency, (2) accelerating the client-side computation, and (3) bringing the content closer to the edge such as directly caching popular content at the 5G base stations.
}

%
%

\begin{figure}[h]
    \centering
    \begin{minipage}{.3\textwidth}
    	\includegraphics[height=.57\textwidth]{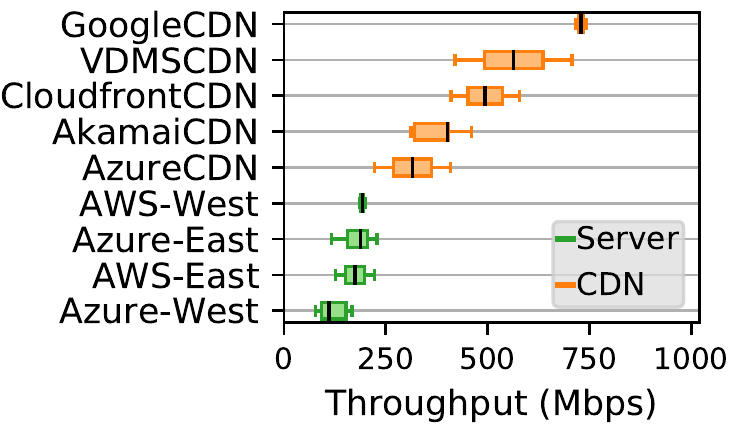}
        \vspace{-.1in}
    \end{minipage}%
    \hfill
    \begin{minipage}{.17\textwidth}
    \includegraphics[height=1\textwidth]{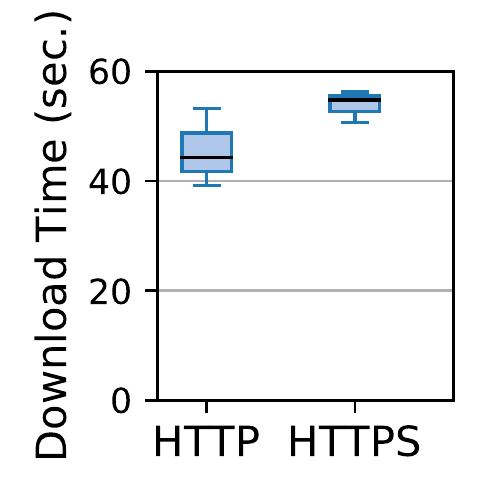}
    \vspace{-.1in}
    \end{minipage}

      \begin{minipage}{.25\textwidth}
  	\caption{\texttt{HTTP(S)} download throughput for 9 CDN/cloud servers over 5G.}
  \vspace{-.22in}
     \label{fig:http}
    \end{minipage}
    \hfill
    \begin{minipage}{.2\textwidth}
  \caption{5G download time (1~GiB file) over \texttt{HTTP} vs. \texttt{HTTPS}.}
    \vspace{-.22in}
    \label{fig:http_vs_https}
    \end{minipage}
\end{figure}


\mysubsection{\texttt{HTTP(S)} Download}
\label{sec:dl}

We investigate the \texttt{HTTP(S)} download performance. We upload a 1~GiB file to  geographically distributed public cloud instances and CDN servers.
We then develop a custom \texttt{HTTP(S)} client that issues 8 parallel byte-range requests each fetching 1/8 of the file over 5G.
The experiments are conducted at Locations A and C (clear LoS with a UE-panel distance of 30m for A and 62m for C, 0\DEG orientation). For each server, we repeat the file download 5 times at both locations, and measure the average throughput.

The results are shown in Figure~\ref{fig:http}. We find that all cloud/CDN servers exhibit low throughput compared to the \hd{iPerf} throughput shown in \feng{Figures~\ref{fig:carrier_8} and~\ref{fig:tcp_th}}: the average throughput ranges from 119 to 730 Mbps with a median of 222 Mbps across all servers. The somewhat surprising results make us realize that \texttt{HTTP(S)} download is very different from \hd{iPerf} bandwidth probing.
Multiple factors may slow down \texttt{HTTP(S)} download, such as the
\texttt{DNS} time,
\texttt{HTTP} request latency,
server-side data processing, 
\texttt{TLS} encryption/decryption,
unbalanced byte-range sessions (some sessions may finish earlier than others~\cite{guo2017accelerating}),
inspection performed by firewall/middleboxes,
to name a few.
Although these factors already exist in 3G/4G eras, they are \emph{amplified} in 5G due to its high speed.
\feng{
Figure~\ref{fig:http_vs_https} exemplifies one such factor: \texttt{TLS} encryption/decryption. In this experiment, we fetch a 1GiB file using \texttt{HTTP/1.1} and \texttt{HTTPS/1.1} from the east coast Azure server running OpenLiteSpeed over mmWave 5G (Locations A and C, clear LoS). To ensure a fair comparison, \texttt{HTTP} and \texttt{HTTPS} downloads are performed back-to-back on the same SGS10 device. As shown, using \texttt{HTTPS} increases the average median download time by 23.5\%, from 44.3 to 54.7 seconds, indicating considerable runtime overhead incurred by \texttt{TLS}.
In contrast, this overhead is barely noticeable in 4G (only $\sim$0.5\% difference between \texttt{HTTP} and \texttt{HTTPS}) due to its lower bandwidth compared to 5G.
}
Overall, the above results indicate that mmWave 5G's high throughput does not always translate to a better application QoE, whose improvement requires joint, cross-layer optimizations from multiple sources.

\mysection{Concluding Remarks}

We conduct a first study that quantitatively reveals 5G performance on COTS smartphones.
In addition to the measurement findings, our study identifies key research directions on improving 5G users' experience in a cross-layer manner.
For example, how to design 5G-friendly transport protocols?
How to strategically select interface(s) among 5G, 4G, and WiFi?
What type of support should a mobile OS provide for enhancing QoE over 5G?
%
Finally, \arvind{we make our dataset \textit{5Gophers}} publicly available to foster the research on 5G.

\begin{acks}
\arvind{We would like to thank the anonymous reviewers for their valuable comments.
We also thank Bo Han for his detailed comments.
This research was in part supported by NSF under Grants CNS-1917424, CNS-1903880, CNS-1915122, CNS-1618339, CNS-1617729, CNS-1814322, CNS-1831140, CNS-1836772, and CNS-1901103.}
\end{acks}

\bibliographystyle{ACM-Reference-Format}
\bibliography{bib/feng,bib/5g,bib/mmwave}

\end{document}